# Laser Spectroscopic Technique for Direct Identification of a Single Virus I:

# FASTER CARS

V. Deckert[1,2,3,] *, T. Deckert-Gaudig[1], D. Cialla[2,3], J. Popp[2,3], R. Zell[4], A. V. Sokolov[1], Z. Yi[1] and M. O. Scully[1,5,6,] *

[1] Institute for Quantum Science and Engineering (IQSE), Texas A&M University, College Station, TX 77843, USA

[2] Institute of Photonic Technology (IPHT) Albert-Einstein-Str. 9, 07745 Jena, Germany

[3] Friedrich-Schiller-Universität Jena, Institute of Physical Chemistry and Abbe Center of Photonics, Helmholtzweg 4, 07743 Jena, Germany

[4] University Hospital Jena, Department of Virology and Antiviral Therapy, Hans-Knöll-Str. 2, 07745 Jena, Germany

[5] Baylor University, Waco, TX 76706, USA

[6] Princeton University, Princeton, NJ 08544, USA

*Corresponding authors: Volker.deckert@uni-jena.de (V. D.), scully@tamu.edu (M.O.S.)

**Abstract**: From the famous 1918 H1N1 influenza to the present COVID-19 pandemic, the need for improved virial detection techniques is all too apparent. The aim of the present paper is to show that identification of individual virus particles in clinical sample materials quickly and reliably is near at hand. First of all, our team has developed techniques for identification of virions based on a modular atomic force microscopy (AFM). Furthermore, Femtosecond Adaptive Spectroscopic Techniques with Enhanced Resolution via Coherent Anti-Stokes Raman Scattering (FASTER CARS) [1] using tip-enhanced techniques markedly improves the sensitivity.

**I. Introduction**

Scanning probe microscopy, especially in combination with plasmon-enhanced near-field spectroscopy, is used to specifically analyze and study objects below Abbe's diffraction limit. The scientific goal is, in addition to virus diagnostics, to identify structural changes of the virus surface at an early stage, using so-called tip enhanced Raman scattering (TERS) together with coherent anti-Stokes Raman spectroscopy (CARS), see Fig. 1. This combined technique will improve the sensitivity and consequently speed up acquisition times considerably. In contrast to known methods, this is unique. Due to its surface specificity our technique allows early detection of change that alter the antigenic properties of viruses and thus the effectiveness of vaccines, with the smallest sample quantities. Furthermore, identification without specific antibodies is possible. This goes far beyond the current state of the art.

In the following section II, we demonstrate how the Tip-Enhanced Raman technique can be used to characterize and differentiate between influenza and picornavirus. In section III we explain the instrumentation and show how it provides exciting new tools for viral research and detection systems, *e.g.* FASTER CARS: Femtosecond Adaptive Spectroscopic Techniques with Enhanced Resolution via Coherent Anti-Stokes Raman Scattering. Section IV presents a conclusion and outlook.

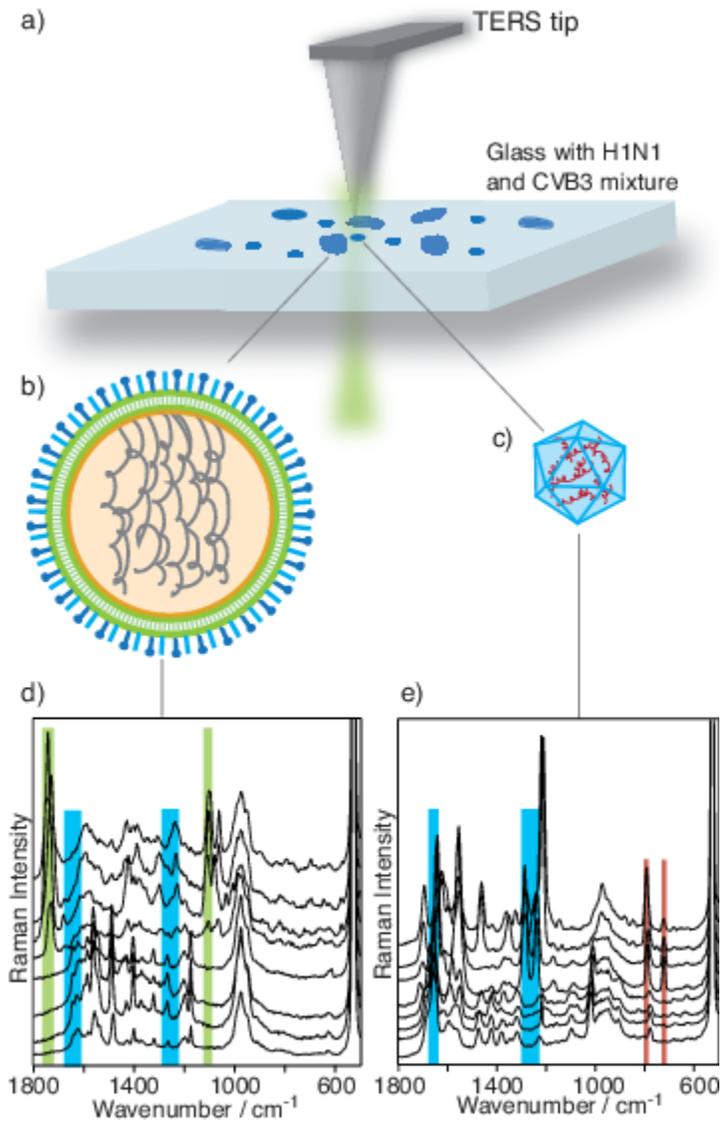

Figure 1: a) Schematics of the transmission TERS setup used in the experiments with the virus mixture spread on a slide and the laser illuminating the sample from below. b) Model of H1N1 virus with protein spikes (blue) protruding from the lipid bilayer (green). c) Model of a CVB3 virus with the protein lattice (blue) on the surface and RNA strands (red) inside the virus. d) Selected TERS spectra recorded on H1N1 virus. e) Selected TERS spectra recorded on CVB2 virus. In d) and e), protein marker bands (blue) and RNA marker bands (red) are highlighted. See the text in section II for further explanation.

## II. Nano-spectroscopic system for differentiating between influenza and picornavirus

A virus is an infectious agent that is spread by nano-sized particles, or virions. As the infection propagates, the virus attacks a cell of a living organism by injecting it with genetic material and inducing the cell to make multiple replicas of the virion. The building blocks of a virion (also called a virus, or a virus particle) are either DNA or RNA molecules surrounded by a protective coat of proteins, the so called (nucleo-)capsid. Moreover, there can be a lipid bilayer serving as envelope. Some viruses are pathogenic and can cause severe diseases, which makes accurate diagnostics essential. The goal is to characterize the specific virus based on single virial particles.  One common technique is the polymerase chain reaction (PCR), which has simplified and accelerated the detection of pathogens over culturing techniques [2-4]. In PCR, the nucleic acid—either DNA or RNA—has to be extracted from the specimen first. Next, characteristic viral sequences are amplified employing varying primer sets followed by further molecular analysis [2,3]. In general, multiple DNA target molecules are necessary for sufficient amplification. The examination of a single virus particle is still challenging. Although PCR technologies are quite powerful, several drawbacks limit the application in microbiological diagnostics. On the one hand, high sensitivity of DNA/RNA amplification makes this process susceptible to contamination that might yield false-positive results. On the other hand, false-negative results, *e.g.* due to failed amplification, have to be considered, too [4]. The nature of the PCR process is prone to several disturbing factors that may hamper exponential DNA amplification. The implementation of the quantitative Real-Time PCR (qRT-PCR) in recent years, where fluorescent molecules are added, enables exact quantitation of template DNA [3,5,6].

Another approach to identifying viruses is the immunoassay-based detection of viral proteins, which is in some cases less sensitive. Thus, the existing molecular analysis tools do not permit a combination of multiplexing and quantitation and in every case the viral components have to be extracted. Working on a single viral particle via the PCR technique is not possible.

Ideally, a new technique would combine the ability of qualitative and quantitative analysis at the single virus particle level, rendering the need for separation of the different components unnecessary. A promising technique for that endeavor is tip-enhanced Raman and potentially other nonlinear optical spectroscopies like FASTER CARS. TERS has been demonstrated to operate in a very specific and sensitive mode down to the single molecule level, with nanometer resolution and below [7-10]. In TERS, scanning probe techniques (atomic force microscopy / AFM or scanning tunneling microscopy / STM) are paired with Raman spectroscopy. The former component enables morphological imaging with sub-nanometer lateral resolution, the latter provides detailed spectral information on every specifically selected position on the sample surface.

The heart piece in TERS is the probe, which generally is a commercial AFM tip (for AFM-based setups) that is commonly evaporated with silver and exactly positioned in the laser focus. Upon laser irradiation of the metallized tip a so-called evanescent field is generated. Molecules located in some nanometer proximity to the tip experience a Raman signal enhancement (up to $10^7$) of their vibrational modes according to the surface-enhanced Raman spectroscopy theory (see for instance refs. [11-15]).

Once the tip is positioned in the laser spot, the sample is moved under the tip and the region of interest is selected. After setting a grid of profile lines on the sample, the AFM is coupled to the Raman spectrometer and enhanced Raman spectra are recorded. Such a setup does not need special sample pretreatments or tagging and allows a direct chemical characterization. The same setup also provides topographic imaging of the sample surface in a single experiment. The spatial resolution in TERS is limited by the diameter of the silver particle at the tip, and lately has been pushed down to around $0.5 - 1$ nm for bio-samples [9, 16, 17]. A not-to-scale setup is sketched in Fig. 1a). For specific instrumental details the reader is referred to ref. [18]. Depending on the number of acquired data, subsequent multivariate data analysis might be useful for data assessment. In the context of a general tip-enhanced approach of a virus identification, it is important to point out that not only all major components to be expected in viruses (namely DNA / RNA [19-21], proteins, lipids and even sugars) have been already detected by TERS. Also, different studies on viruses [22-24] already proved the feasibility of the concept. In contrast to PCR-related techniques, in TERS a separation of the different components or labeling is not required, and ultimately only a single virion is sufficient for an identification.

So far, reported TERS experiments on single virus particles were performed on more or less randomly chosen points, which is apparently not sufficient for a thorough characterization [22-24]. It was shown that grid-based TERS mapping of *Varicella-zoster* and *Porcine teschovirus* allowed a linear discriminant supported distinction of the two viruses. Clearly, the identification and discrimination of different virus strains demand a comprehensive characterization in terms of spectral surface imaging which is in line with the above mentioned extreme lateral resolution.

For the present experiments an enveloped influenza A virus H1N1 and a non-enveloped coxsackie virus B3 (CVB3) were chosen. The swine H1N1 influenza virus particle is composed of a 2-6 nm lipid bilayer decorated with viral proteins and eight ribonucleoprotein (RNP) complexes. The RNPs consist of RNA strands, which interact with numerous nucleoprotein molecules and some viral polymerase complexes. The consistency of the lipid envelope depends on the type of cell membrane of host cell the virus originates. The inner surface is covered completely with viral matrix protein. From the lipid layer flexible protein spikes with a protrusion height of 10-14 nm stick out [25]. In Fig. 1b) the structure of an influenza virus is sketched. From AFM measurements it is known that H1N1 shows pleomorphism that means size $(150 - 400 \text{ nm})$ and shape (spherical or rod-like) vary [25, 26]. The dimension of the CVB3 virus (Fig. 1c)) is ten times smaller (20-30 nm), with an icosahedral structure (determined by crystallography), where the RNA is packed in a protein coat (capsid) [27]. Clearly, the different surface components of the viruses yield TERS spectra with information on proteins and lipids for H1N1 (Fig. 1d)) and protein and RNA for CVB3 (Fig. 1e)). Again, it is obvious that spectral variability on the virus surface occurs and a larger virus surface area must be considered. Nevertheless, a discrimination of both viruses solely based on the TERS spectra is quite straightforward. The experiment with the full statistical evaluation (multiple maps, different particles, different tips) will be published elsewhere. Here we want to emphasize that the method is potentially able to identify any single virus based on the surface composition. The main challenge is the time required for an assessment, particularly the spectral acquisition time. Interestingly, the necessary AFM topography which is always detected to locate a sample can also provide substantial information for a virus pre-screening, thus considerably limiting the potential candidates for TERS investigations. Another possibility to increase the sensitivity will be introduced in the following section where we combined tip/surface enhanced Raman techniques with FAST CARS.

## III. Instrumentation and new techniques

To our best knowledge, tip-enhanced CARS has been demonstrated in early 2000s [28]. Following our success in detecting anthrax using FAST CARS [29], we have developed new vibrational spectroscopic techniques for nanoscale real-time molecular sensing having large signal enhancement, small background, short detection time and high spectral resolution [30]. On other hand, other groups have pushed the surface-enhanced CARS to single molecule sensitivity [31, 32]. Our time-resolved Tip Enhanced Coherent Anti-Stokes Raman Spectroscopy (TECARS) is an exciting new tool for virus research. Our early "proof of principle" research was used to detect hydrogen-bonded molecular complexes of pyridine with water in the near field of gold nanoparticles with large signal enhancement. This yields an improved FAST CARS [1] approach which is aptly called FASTER CARS.

Combining this technique with modern quantum molecular calculation is natural for studying complex biological systems. Fig. 2 shows a simplified experimental arrangement used in ref. [30]. Fig. 3 shows typical data.

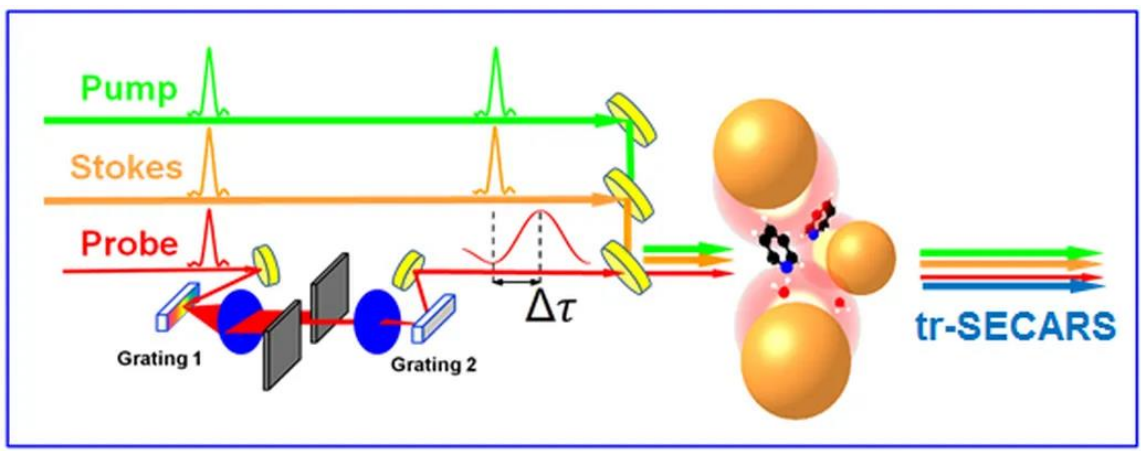

Figure 2: Simplified experimental setup of time-resolved surface-enhanced CARS spectroscopy. Three laser beams collinearly excite nanomolar amounts of pyridine in the near-field of gold nanoparticles. The pump (green) and Stokes (orange) broadband femtosecond laser pulses excite a molecular vibrational coherence which is probed by a time-delayed shaped narrowband picosecond probe pulse (red). The probe pulse is Sinc-shaped by a pulse shaper slit. The CARS signal is generated in the forward direction (blue) and collected by a spectrometer with a 0.1 s integration time.

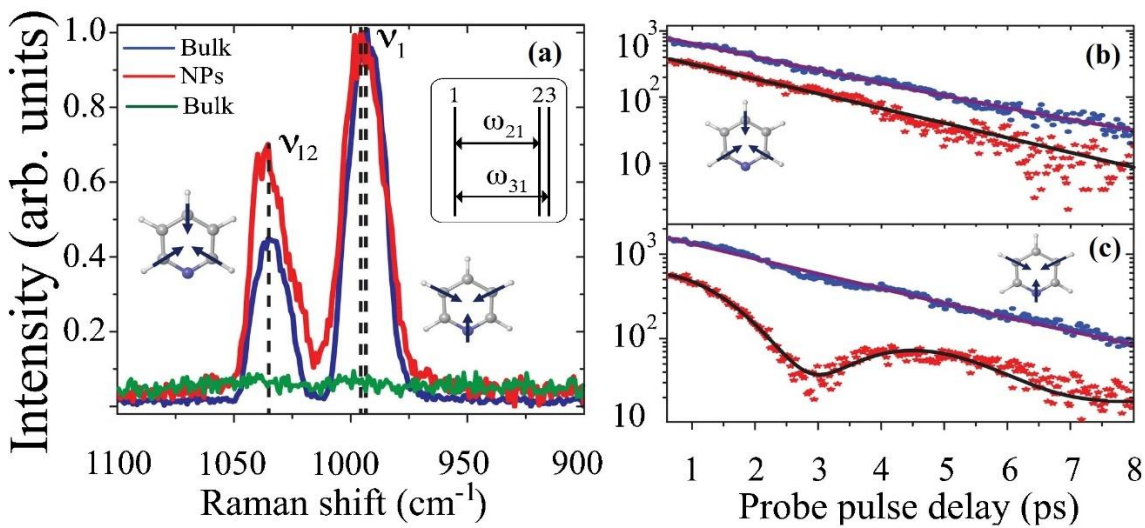

Figure 3: Normalized spectra for 1 ps delay and temporal traces of pyridine with (red) and without (blue, green) gold nanoparticles (NPs). Signals with NPs were obtained from 12 µm thick layers of pyridine on the surface of NPs. Bulk signals without NPs showed no detectable spectral features for 12 µm thick samples (green, (a)). Therefore, bulk signals without NPs were obtained under similar excitation conditions from 2 mm thick layers (blue). The ring breathing mode of the pyridine-water complex is not resolved in the spectra (a), but can be extracted from the temporal trace (c) where the vibrational dephasing is deconvoluted from the probe pulse shape. No detectable signals were observed without NPs under identical conditions (green) and, therefore, the amount of bulk pyridine was increased (blue). Additional details can be found in ref. [30].

The review article of Lis and Ceccht characterizes this technique nicely; they say [33]:

"Investigations demonstrated that the electronic background that derives from the water and the metal could also be reduced on such kind of solid substrates.

Voronine *et al.* used time-resolved surface-enhanced CARS, which combines delayed laser pulses of different widths, to obtain a high spectral resolution with the suppression of the non-resonant back ground."

They go on to say that this technique yields "astonishing sensitivity". Indeed, this improvement over our earlier FAST CARS anthrax detection scheme holds real promise for detection and identification of single virus particles such as SARS-COV-2 virus which causes COVID-19.

**IV Conclusion and outlook**

We have shown that on one hand, the Tip-Enhanced Raman technique can be used to characterize and differentiate between influenza and picornavirus. On the other hand, surface enhanced CARS result in "astonishing" enhancement in the sensitivity [33]. The new FASTER CARS system, which combines nanometer spatial resolution and enhanced sensitivity, forged by both the tip and coherent Raman enhancement, provides exciting new tools for viral research and detection, e.g., the SARS-COV-2 virus.

**Acknowledgements:**

We would like to express our deep appreciation to TAMUS chancellor John Sharp without whom this project would not have happened. We thank the: Office of Naval Research (Award No. N00014-16-1-3054), Air Force Office of Scientific Research (Award No. FA9550-18-1-0141), National Science Foundation (Awards No. 1828416, 1609608), Chancellor's Research Initiative/Governor's University Research Initiative (CRI/GURI), the Robert A. Welch Foundation (Grant No. A-1261), King Abdulaziz City for Science and Technology (KACST) and the Deutsche Forschungsgemeinschaft (DFG, German Research Foundation, Grant No. DFG-CRC1375NOA) for support of the work. We also like to thank Drs. G. Agarwal, B. Brick, K. Chamakura, J. Mogford, J. Sharp, R. Young, and A. Zheltikov for stimulating and helpful discussions.